\documentclass[floatfix,twocolumn,rmp,superscriptaddress,showpacs]{revtex4}

\usepackage{dcolumn,graphicx,amsmath,amssymb}
\usepackage{txfonts}

\begin{document}

\title{Network reachability of real-world contact sequences}

\author{Petter Holme}
\affiliation{Department of Physics, University of Michigan, Ann Arbor,
MI 48109}

\begin{abstract}
  We use real-world contact sequences, time-ordered lists of contacts
  from one person to another, to study how fast information or
  disease can spread across network of contacts. Specifically we
  measure the \textit{reachability time}---the average shortest time
  for a series of contacts to spread information between a reachable
  pair of vertices (a pair where a chain of
  contacts exists leading  from one person to the other)---and the
  \textit{reachability ratio}---the fraction of reachable vertex
  pairs. These measures are studied
  using conditional uniform graph tests. We conclude, among other
  things, that the network reachability depends much on a core where
  the path lengths are short and communication frequent, that
  clustering of the contacts of an edge in time tend to decrease the
  reachability, and that the order of the contacts really do make
  sense for dynamical spreading processes.
\end{abstract}


\maketitle

\section{Introduction}

\begin{figure}
  \resizebox*{\linewidth}{!}{\includegraphics{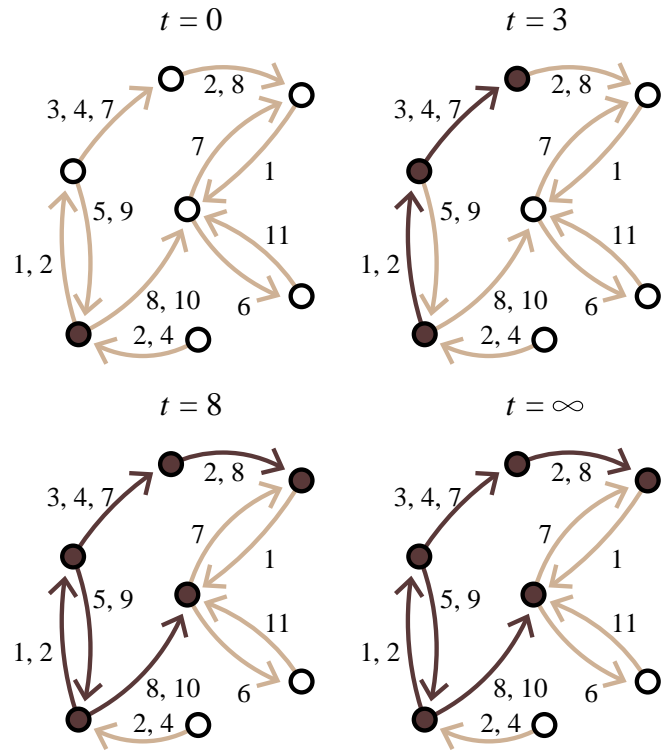}}
  \caption{
    An illustration of optimal spreading processes through a contact
    sequence. The information transfer (the contacts) occurs in the
    direction of the arrows at the times marked on the edges. The vertices
    that can have the information, and the time respecting paths, are
    shown in darker color.
  }
  \label{fig:ill}
\end{figure}

The advent of modern database technology has greatly vitalized the
statistical study of networks. The vastness of the available data sets
makes this field apt for the techniques of statistical
physics~\cite{ba:rev,mejn:rev,doromen:book}. One
particular example that has been extensively studied is the contact
networks of individuals~\cite{pok,nioki,bornholdt:email,gui:sesi,mejn:liame}.
The vertices in this kind of networks are individuals and an edge
between two people means that there has been a contact between these
persons. Typical data sets for this kind of networks are lists of
messages through some electronic medium, like
e-mails~\cite{bornholdt:email,gui:sesi,mejn:liame} or instant messenger
applications~\cite{nioki}. In many cases, the times of the contacts may
also be available, which makes the data set much more informative than
the corresponding contact network. Some studies of the temporal
statistics of such data sets have been
made~\cite{eckmann:dialog,my:ongoing,johansen:resptimes}, but
how the contact dynamics affect the picture from the network topology is yet,
in many respects, obscure. In this paper we study some aspects of the
temporal contact pattern in a the framework of networks: How fast can
information, or disease, spread across an empirical contact network?
How much of the network can be reached from one vertex through a
series of contacts? These are
properties that depends not only on the number of neighbors of a
vertex, or the number of contacts along an edge, but also the time
ordering of the contacts~\cite{riolo:contact,tempo}. In
Fig.~\ref{fig:ill} we give an illustration of such spreading
processes.

We call a network
where information can spread between most pairs of vertices through a series of
contacts, and where the spreading is comparatively fast, a network
with high \textit{reachability}. We quantify and measure reachability for four
data sets from e-mail exchange and contacts within an Internet
community. The results are analyzed by a systematic use of conditional
uniform graph tests. The reachability scaling in the
large-time limit is also discussed. We use the metaphor of information, or rumor,
spreading, but the results applies to a range of dynamical processes.

\section{Preliminaries}

\subsection{Definitions}

By a \textit{contact sequence} we mean a set of ordered triples
$C=\{(i,j,t)\}$, where the triple is referred to as a \textit{contact}
from the vertex (person) $i$ to the vertex $j$ at time $t$ (or along the edge
$(i,j)$ at time $t$). For simplicity we set the first time of
$C$ to zero, so the times of $C$ belongs to the interval $[0,t_\mathrm{stop}]$. We let $L$ denote number of contacts and $M$ the number
of directed edges. The (contact)
network is a set of $N$ vertices $V$, together with
a set of $M$ directed edges $E$ such that a pair $(i,j)$ of vertices is
a member of $E$ if and only if there is a time $t$ such that
$(i,j,t)\in C$. We let $k$ denote the degree of a vertex (the number
adjacent edges). Since the edges are directed it is sensible to distinguish
between in- and out-degree. We let $l$ denote the number of contacts
along a particular edge. Following Ref.~\cite{tempo} we call a list of contacts
of increasing times a \textit{time respecting path}. Just ``path''
will refer to a path, in general, in the contact network. We note that no
other sets of contacts, other than time respecting paths, can transfer
information, disease or commodity from one vertex to another in a
contact sequence. We let
$\tau(i,j,t)$ denote the shortest time to reach $j$ starting from $i$
at time $t$ and  $\tau(i,j)$ denote the corresponding time-averaged
quantity. One problem is that there is not always a time-respecting
path from one vertex to another. We deal with this by both looking at
the \textit{reachability time} $\hat{\tau}$---$\tau$ averaged over all
pairs such that there exist a time-respecting path connecting them,
and \textit{reachability ratio} $f$---the fraction of the vertex-pairs
that does have a time respecting path between them. A way to
characterize the network reachability with only one number would be to
consider the harmonic mean of $\tau$ (which is well-defined even if
there are unreachable vertex-pairs). We do not do that since it is
not a very intuitive quantity in that it is not the time-average  of
some actual process in the system.

\subsection{The data sets}

We use four real-world contact sequences all derived from
communication over the Internet. One of the data sets is based on
contacts within the Swedish Internet community pussokram.com primarily
intended for romantic interaction.~\cite{pok} In this data set the
edges can represent messages of e-mail type, but they can also
represent writing in guest books that are visible to the rest of the
community. The other three contact sequences are complied from e-mail
exchange. The data set of Ebel \textit{et al.}~\cite{bornholdt:email}
and Eckmann \textit{et al.}~\cite{eckmann:dialog} are constructed from
log files of e-mail servers at two universities.
In the Ebel \textit{et al.} data set a least one
vertex of each contact is a student or employee of the university. The
network of the outer vertices (the ones not corresponding to a e-mail
account hosted by the university) is not mapped. For the Eckmann
\textit{et al.}\ data no outer vertices are present.
The fourth data set is constructed
from the e-mail directories of 150 top executives of the Enron
corporation. This data set was released to public during the legal
investigation concerning the Enron
corporation. (Available at http://www-2.cs.cmu.edu/~enron/.) The
contact sequence was constructed by parsing the e-mail headers and
adding contacts from $i$ to $j$ if the address $i$ appears in the ``From''
field of the same e-mail where the address $j$ is present in the ``To''
field (we do not include addresses in the ``Cc'' and ``Bcc''
fields). As a result this data set contains contacts between outer
nodes, if the e-mail is from an outer vertex $i$ and some other
recipient $j$ is also an outer vertex then there will be a contact
from $i$ to $j$ in the contact sequence. The disadvantage of the Enron
data set, compared with the other e-mail data sets, is that some of
the e-mails that really were sent to, or received by, the persons has been
deleted, either by the individuals themselves or during the
preparation of the data set for the sake of protecting privacy. The
sizes of the data sets can be found in Table.~\ref{tab:size}. The time
resolution is one second for all data-sets.

\begin{figure}
  \resizebox*{\linewidth}{!}{\includegraphics{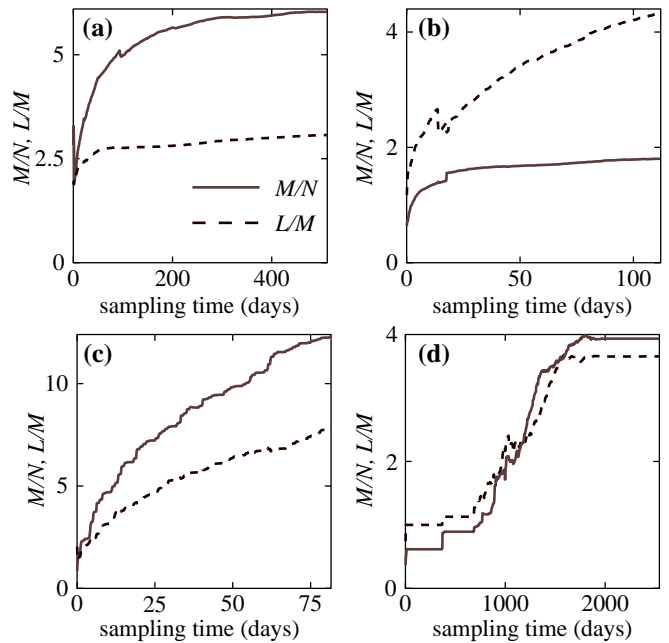}}
  \caption{
    The average number of edges per vertex (or the average in- or
    out-degree) and the average number of contacts per edge for our four
    data sets: (a) the Swedish Internet dating community
    pussokram.com, (b) the e-mail data of Ebel \textit{et al.}, (c) the
    e-mail data of Eckmann \textit{et al.}, and (d) the e-mail data of Enron
    executives.
  }
  \label{fig:ncl}
\end{figure}

To get a feeling for the data, we plot the average number of directed
edges per vertex and the average number of contacts per directed edge
in Fig.~\ref{fig:ncl}. We note that the four data sets differ much,
both in the actual values of $M/N$ and $L/M$ and the shape of the
curves. We note that the Enron curves in Fig.~\ref{fig:ncl}(d) are
very flat for early and late sampling times, this is because quite few
mails are dated to these intervals. It would be reasonable to
preprocess the data by discarding the early and late Enron e-mails, or
removing the spam-mails causing the jagged appearance of the curves in
Fig.~\ref{fig:ncl}(b), but this would require a more detailed
knowledge of the data. Instead we simulate the data sets as they are
and assume that the conclusions will be qualitatively correct due to
the much larger amounts of good data. (Some effects of strange
communication will be discussed explicitly below.) From
Fig.~\ref{fig:ncl} we can also conclude that all
properties of contact sequences from computer mediated communication
are not general---different settings of the data collection can
record different contact patterns. This is nothing special for this
kind of data, but it indicates that generalizations must be made with caution.

\subsection{Numerical procedures}

The size of the data sets has consequences for how e.g.\ $\hat{\tau}$ is
to be calculated in practice. If the data set is small, one can compute
$\tau(i,j,t)$ for each vertex-pair an every time occurring in
$C$---for any $t$ is not in $C$ we have
\begin{equation}
  \tau(i,j,t)=\tau(i,j,t')+t-t'
\end{equation}
where $t'$ is the largest time in $C$ that is smaller than $t$. Our
data sets are too large for such a procedure (at the time of
writing). Instead we sample 100 times randomly over a
interval $[0,\nu t_\mathrm{tot}]$ and calculate $\tau(i,j,t)$
for each vertex-pair. To use a $\nu$ less than one will give longer
time respecting paths (as time respecting paths originating from
individuals only present in the end of the
data set are omitted). Since the fraction of $O(N)$-paths will
decrease with the sampling time, choice of $\nu$ matters less the larger the
contact sequence is, and the choice of $\nu$ does not alter any
qualitative conclusions. We use $\nu=0.3$ throughout the study. The
actual calculation of $\tau(i,j,t)$
can be done in $O(L)$ time by initially marking $i$ by $t$ and every other vertex
``unvisited,'' then running through $C$ in the order of
increasing times and for every triple $(i,j,t')$ marking $j$ with $t'$
if and only if $j$ is marked ``unvisited'' and $i$ is marked with a
time tag.

\subsection{Conditional uniform graph tests}

To put the observed $\hat{\tau}$ into perspective, and understand how it
results from the temporal contact pattern and the network topology, we
compare the measured values with values averaged over ensembles of
randomized contact sequences in ``conditional uniform
tests.''~\cite{katz:cug}. By systematically integrating (or rather,
averaging) out different types of structure one can see how these types of
structure are contributing to the measured result. For example, if we
want to assess the impact of the order of the contacts we compare
$\tau$ of $C$ with $\tau$ averaged over contact sequences with
the times of $C$ randomly permuted. The five conditional uniform test
ensembles we use, all having the same number of vertices, edges and
number of contacts as $C$, are the following: 
\begin{description}
\item[Permuted Times (PT)] The set of contact sequences with the times
  randomly permuted, i.e.\ the edges, and the number of occurrences of
  a particular edge in $C$ is unchanged, as is the set of times, only
  the times of the communication along a particular edge is randomized.
\item[Random Times (RT)]
  The ensemble with the same edges, and the same number of them as $C$,
  but the times are chosen uniformly randomly in the same interval as the
  times of $C$.
\item[Random Contacts (RC)]
  With the same set of edges as $C$, but the numbers of contacts per edges are
  chosen at random (so it can be zero). The times are chosen at random
  in $C$'s time span. 
\item[Randomized Edges (RE)] The ensemble of contact sequences with the same
  set of degrees as the network generated by $C$. The time list is the
  same as $C$, so that the $n$'th contact of the randomized sequence represents
  a contact between vertices of the same degrees as in $C$ and occurs
  at the same time as the $n$'th contact of $C$.
\item[All Random (AR)] The contacts are completely random, as are
  the times, only the sizes $L$, $M$, $N$ and $t_\mathrm{tot}$ are common to the
  original network. Thus the corresponding graph is a Poisson random
  graph~\cite{janson}.
\end{description}
We note that the principles behind conditional uniform test are rather
similar to those behind exponential random graph
models.~\cite{strauss}. Such models, as usual in statistical physics,
are based on the assumption that when the constraints of a model are
much fewer than the degrees of freedom, the best choice of model is the
one that maximizes the entropy.~\cite{park:sm} The difference is that
conditional uniform graph tests usually have $O(N)$ constraints,
whereas exponential random graphs typically have $O(1)$ constraints.

We sample 100 randomized contact sequences for each conditional
uniform graph test.

\begin{table}
\caption{\label{tab:size} The number of vertices $N$, edges $M$ and
  contacts $L$ as well as the total sampling time $t_\mathrm{tot}$ for
  our four data sets.}
\begin{ruledtabular}
\begin{tabular}{r|llll}
 & $N$ & $M$ & $L$ & $t_\mathrm{tot}$ (days) \\\hline
pussokram.com & 29341 & 174662 & 536276 & 512.0 \\
Ebel \textit{et al.} & 57194 & 103083 & 447543 & 112.0 \\
Eckmann \textit{et al.} & 3188 & 39256 & 309125 & 81.7 \\
Enron & 78592 & 308147 & 1119874 & 2551.0 \\
\end{tabular}
\end{ruledtabular}
\end{table}

\begin{table*}
\caption{\label{tab:res} The reachability time $\hat{\tau}$ in days
  and the reachability ratio $f$
  for the different networks. The one s.d.\ errors are of the order of
  the fourth non-zero digits. $\hat{\tau}$ is plotted against $f$ in
  Fig.~\ref{fig:scp}.
}
\begin{ruledtabular}
\begin{tabular}{r|llllll|llllll}
 & \multicolumn{6}{c}{reachability time $\hat{\tau}$} &
 \multicolumn{6}{c}{reachability ratio $f$} \\
& real & PT & RT & RC & RE & AR& real & PT & RT & RC & RE & AR\\
\hline
pussokram.com & 219 & 173 & 213 & 192 & 233 & 316 & 0.289 & 0.471 &
0.512 & 0.597 & 0.236 & 0.878 \\
Ebel \textit{et al.} & 74.7 & 71.7 & 70.1 & 61.6 & 70.4 & 83.1 &
0.0199 & 0.0380 & 0.0381 & 0.123 &0.0162 & 0.00504 \\
Eckmann \textit{et al.} & 22.8 & 21.0 & 20.5 & 10.1 & 23.9 & 7.61 &
0.538 & 0.593 & 0.592 & 0.657 & 0.545 & 1.00 \\
Enron & 1274 & 1196 & 1103 & 926 & 1290 & 1932 & 0.0544 & 0.0839 &
0.0724 & 0.0973 & 0.0596 & 0.338\\
\end{tabular}
\end{ruledtabular}
\end{table*}

\section{Results}

In this section we give, using conditional uniform tests, a thorough
discussion about what structures of the data that govern the
reachability. After that follows two short sections on the how
the reachability is influenced by the traits of the start and finish
vertex of a time-respecting path, and extrapolation of the results to the
limit of large times.

\subsection{The reachability times and reachability ratio}

\begin{figure}
  \resizebox*{\linewidth}{!}{\includegraphics{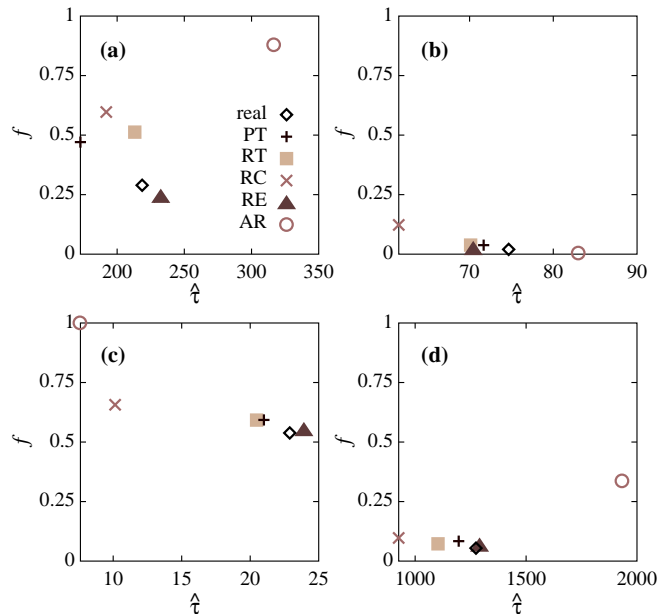}}
  \caption{
    The fraction of reachable vertex-pairs (vertex-pairs between which
  a time respecting path exist) plotted against the average
  reachability time for reachable vertex-pairs. (a) shows results for
  the pussokram.com data, (b) represents the e-mail data of Ebel
  \textit{et al.}, (c) is the results of the of Eckmann \textit{et
  al.}\ e-mail data, and (d) the e-mail data of Enron executives. The
  actual values are tablelized in Table~\ref{tab:res}.
  }
  \label{fig:scp}
\end{figure}

The values for $\hat{\tau}$ and $f$ are given in Table~\ref{tab:res}
and plotted against each other in Fig.~\ref{fig:scp}. Good network
reachability means low values of $\hat{\tau}$ and high values of
$f$. Thus points in the the upper left corner of the plots in
Fig.~\ref{fig:scp} represent high reachability, while
the lower right corner mean low reachability. Which of
$\hat{\tau}$ and $f$ that is the more important may vary: If
one is interested of how soon information can reach a certain small
number of people then $\hat{\tau}$ is the more relevant. If
one is interested of how many that eventually will receive the
information, then $\hat{\tau}$ is the more relevant.

The actual values of $\hat{\tau}$ and $f$ for the real-world data present few surprises. The largest
$f$ is observed for the densest Eckmann \textit{et al.}\ e-mail data,
which is natural since the size of the largest connected component is
known to increase with average degree. We also note that the sparsest
data (of Ebel \textit{et
  al.})\  has the smallest $f$. That the largest dataset, the
Enron data, has the largest $\hat{\tau}$ and that the smallest data
set (of Eckmann \textit{et al.})\  has the smallest $\hat{\tau}$ is of
course no surprise either. 

For the conditional uniform tests we begin comparing the
original networks with the one randomized according to the PT
constraints. PT, only involving permuting the times of the contacts,
is the test giving the smallest change to the original contact sequences. The
reachability ratio of the rewired networks are consistently larger and the
reachability time smaller. A natural explanation for this is that
people tend to engage in dialog with
each other \cite{eckmann:dialog,my:ongoing,johansen:resptimes} for a
while, thus the contacts along one edge tend to be clustered in
time. For the reachability this means that the first contact of such a
dialog will carry most of the time respecting paths. If the contacts
are more evenly spread out in time (the effect of the PT randomization)
the waiting time between the contacts will decrease and so will the
reachability time. If the message exchange along particular edges would not be clustered
in time the spread of information would thus
probably be much faster. One may think of real world processes that
increases the reachability---there may exist waves of large-scale
message relaying, but apparently the dialog effect overshadows the
impact of such events. Already from this comparison we see that one
looses much information from the contact sequence if one only consider
the network it defines, no matter if the network is
weighted~\cite{bar:wei} or not.

The RT test generates slightly more randomized contact sequences than
PT. In this case the times do not follow the daily routines of
people---the probability of response after a certain time has peaks at
multiples of days (for e.g.\ the data in Ref.~\cite{bornholdt:email}
there is even a larger peak after one weak) reflecting peoples' daily
(and weekly) routines~\cite{my:ongoing,johansen:resptimes,tyta:resp,bego:wory}. As seen in
Table~\ref{tab:res} the differences between the PT and RT
randomizations are not very big. This is maybe
not very surprising since the sampling time is much larger than the
mentioned daily and weekly routines (that gives a structure that is
removed by RT, and could make a difference to PT). A structural bias that
would favor reachability in the RT randomized sequences is that, in the
real data, some edges, and vertices are created or ceased
(cf.\ Ref.~\cite{my:ongoing}) during the sampling procedure which makes
them inaccessible, in a time respecting sense, for early or late times
respectively. Evidently, this has no major impact on the reachability.

\begin{figure}
  \resizebox*{\linewidth}{!}{\includegraphics{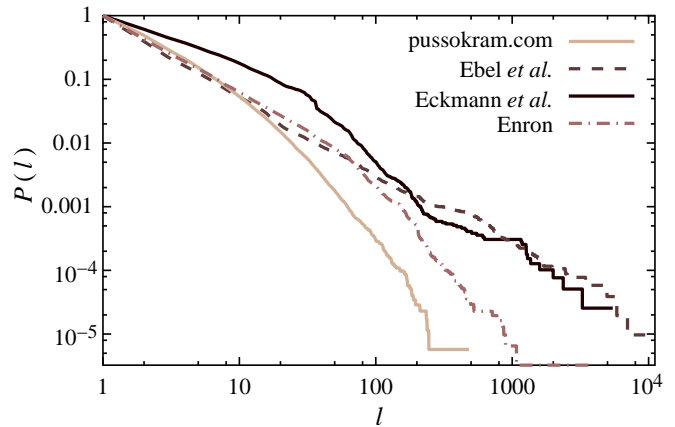}}
  \caption{ The cumulative distribution of the number of contacts over
    a directed edge $l$ (i.e., the probability that an observed $l$ is
    equal to or larger than the value on the abscissa). All data sets
    show right-skewed and broad distributions of $l$.
  }
  \label{fig:cldist}
\end{figure}

The RC randomization outputs contact sequences with yet less
structural constraints
than the RT scheme. As it turns out---see Fig.~\ref{fig:cldist}---the
distribution of contacts along a particular edge $l$ is skewed for
all the data sets, with broad tails. This behavior was observed for
scientific collaboration and airline networks in Ref.~\cite{bar:wei},
models of this behavior are given in Refs.\ \cite{goh:wei} and \cite{ves:wei}. These broad
tails are replaced by a Poisson $l$-distribution 
for the RC randomized networks. The more frequent contacts along the
edges having few contacts in reality increases the network
reachability, both in terms of the reachability ratio and reachability
time (all RC points in Fig.~\ref{fig:scp} lie above and to the left of
the corresponding RT points). A closer look at Fig.~\ref{fig:cldist}
shows that the largest $l$ values for some of the data sets are
probably unrealistically big for a communication in the normal
sense. Some identical contacts are present in the Ebel \textit{et
  al.}\ data sets (i.e.\ e-mails with the same sender and recipient
sent the same second) these possibly comes from the same e-mail being
addressed to the same receiver multiple times). If such multiple
contacts were filtered away, the reachability would be unaltered for
the real networks, but larger for the randomized networks (since they
would be denser in contacts). One can also argue that such events
should be retained in some cases as spam (and other one-to-many
messages), in theory, can be read and spread information. We conclude
that one needs
to keep the presence of such abnormal edges in mind in the evaluation
of contacts sequences.

\begin{table}
\caption{\label{tab:llc} The correlation coefficient $r_{ll}$ for the
  number of contacts for adjacent edges---edges forming a directed
  path of length two. $r_{ll}^\mathrm{RE}$ gives the reference value
  for ensembles of networks with the same degree sequence as the
  original. $r_{ll}$ is consistently bigger than $r_{ll}^\mathrm{RE}$
  for all networks. The digit in parentheses gives the one s.d.\ error
  of the last digit. 50 RE randomizations are used.
}
\begin{ruledtabular}
\begin{tabular}{r|dd}
 & r_{ll} & r_{ll}^\mathrm{RE} \\\hline
pussokram.com & 0.0656 & 0.0003(3) \\
Ebel \textit{et al.} & 0.00188 & 0.0001(3) \\
Eckmann \textit{et al.} & 0.00626 & -0.0003(7) \\
Enron & 0.0251 & -0.0007(2)\\
\end{tabular}
\end{ruledtabular}
\end{table}

\begin{figure*}
  \resizebox*{!}{6.5cm}{\includegraphics{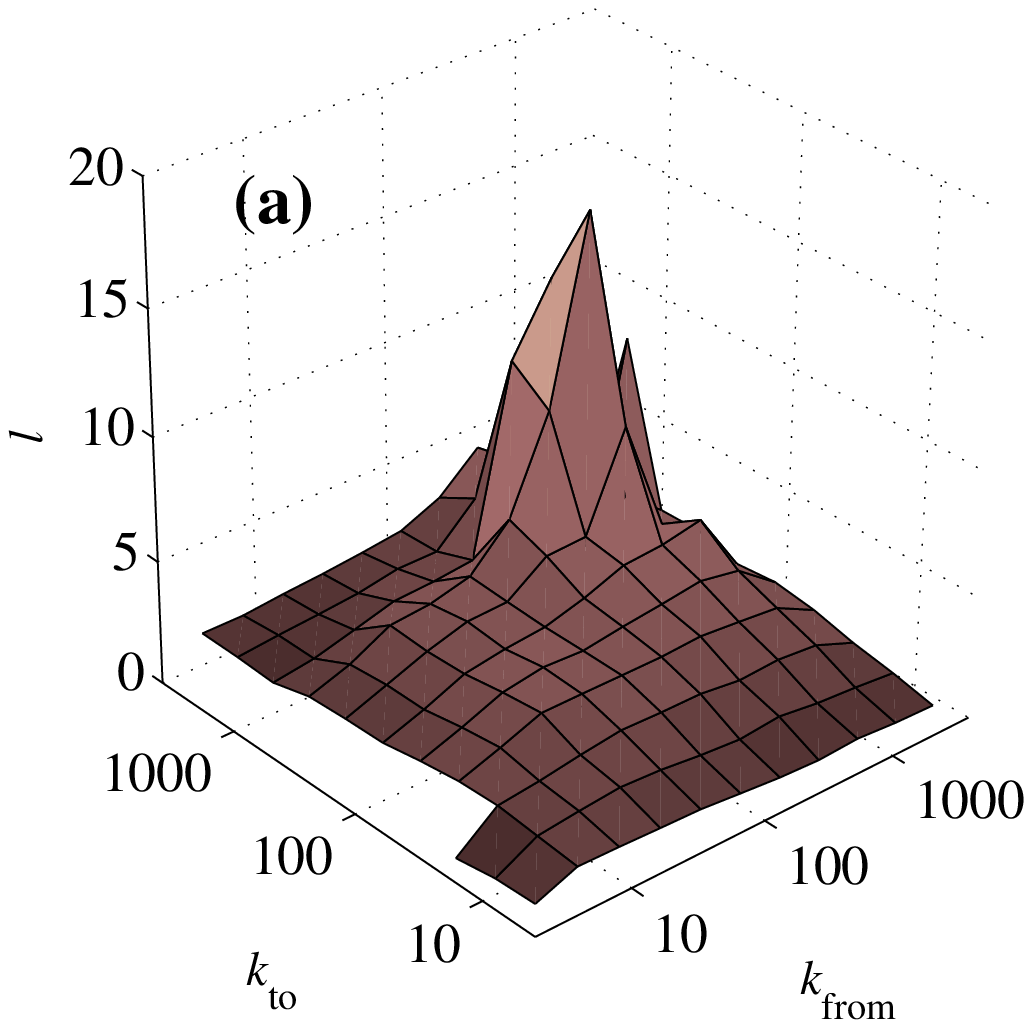}}
  \resizebox*{!}{6.5cm}{\includegraphics{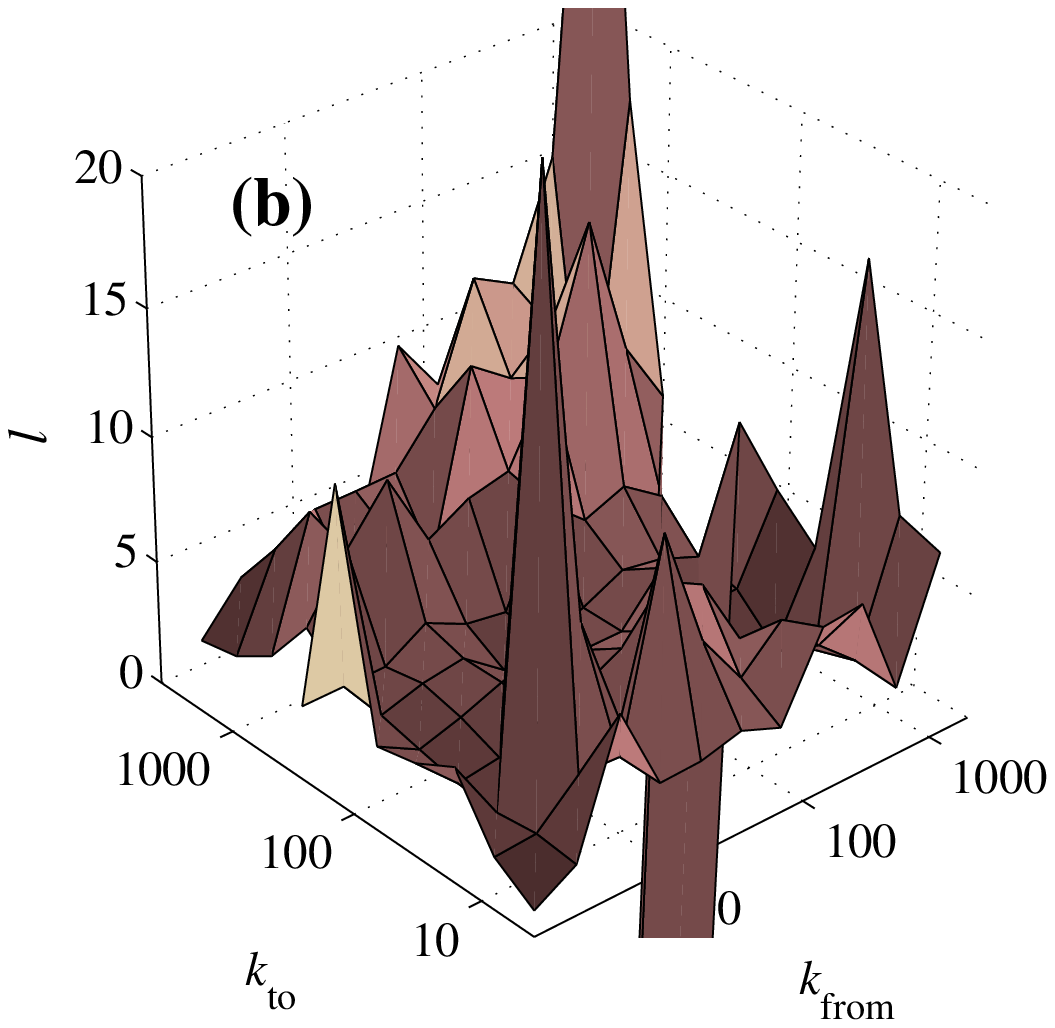}}\\
  \resizebox*{!}{6.5cm}{\includegraphics{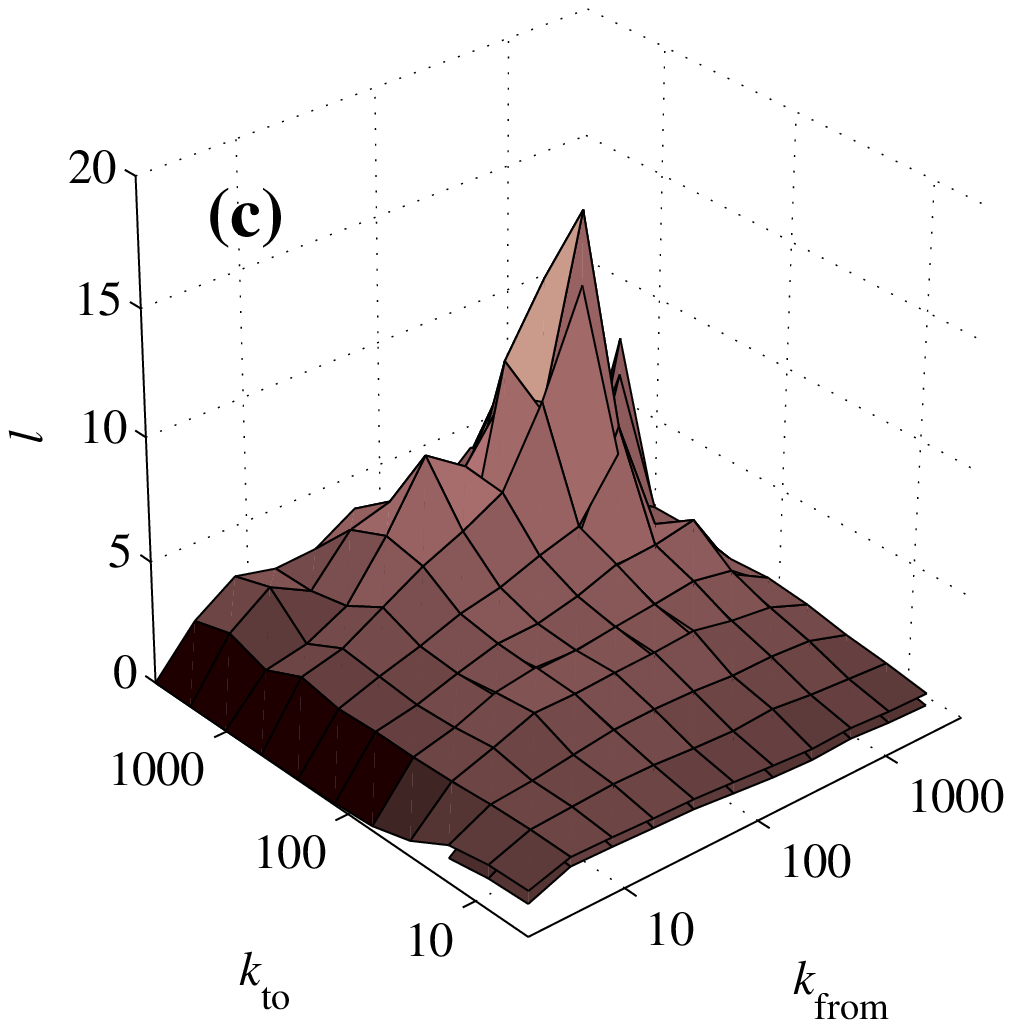}}
  \resizebox*{!}{6.5cm}{\includegraphics{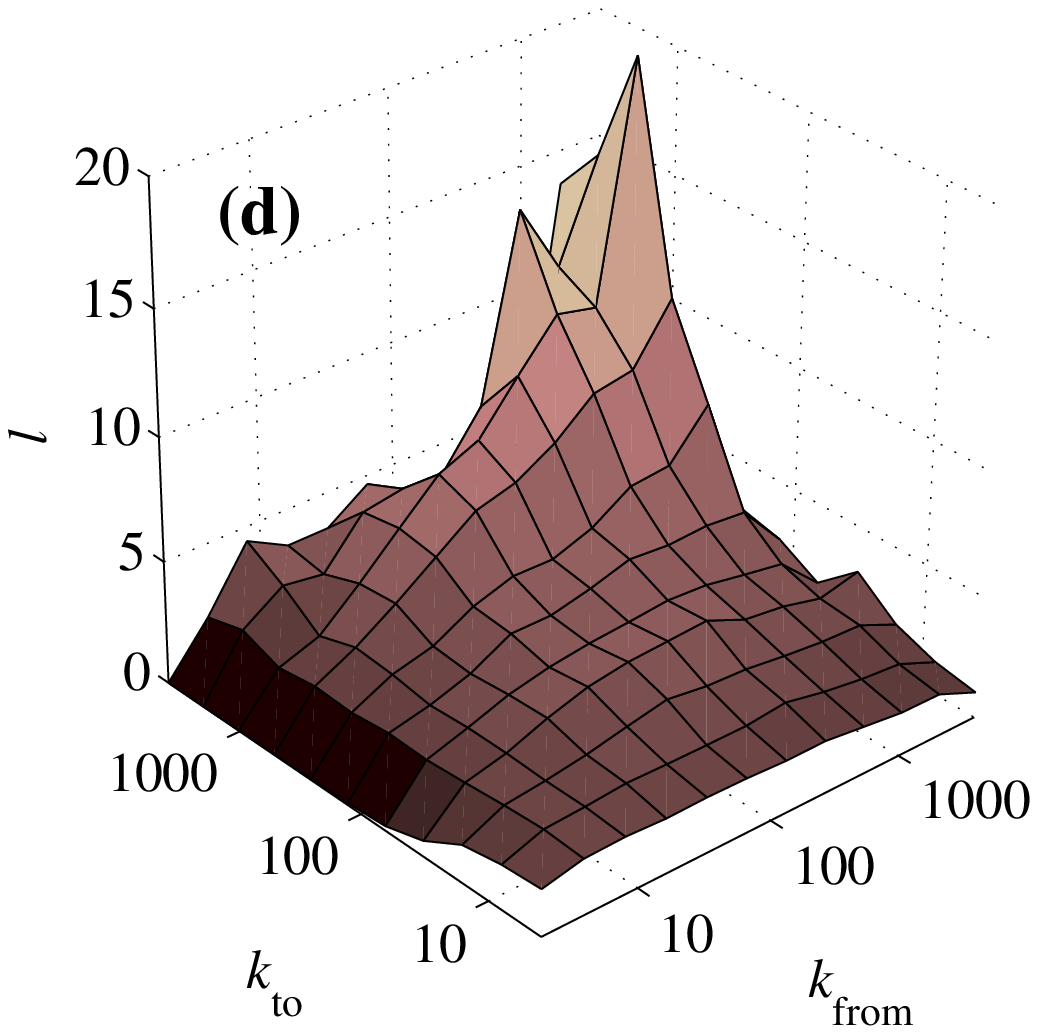}}
  \caption{ The number of contact per edge $l$ plotted against the
    out-degree of the from-vertex (i.e.\ the first argument of the
    directed edge) $k_\mathrm{from}$ and the in-degree of the
    to-vertex $k_\mathrm{from}$. The plots are for the (a)
    pussokram.com, (b)  Ebel \textit{et al.}, (c) Eckmann
    \textit{et al.}, and (d) Enron data respectively. The abscissae
    are logarithmically binned.
  }
  \label{fig:lkk}
\end{figure*}

\begin{figure*}
  \resizebox*{!}{6.5cm}{\includegraphics{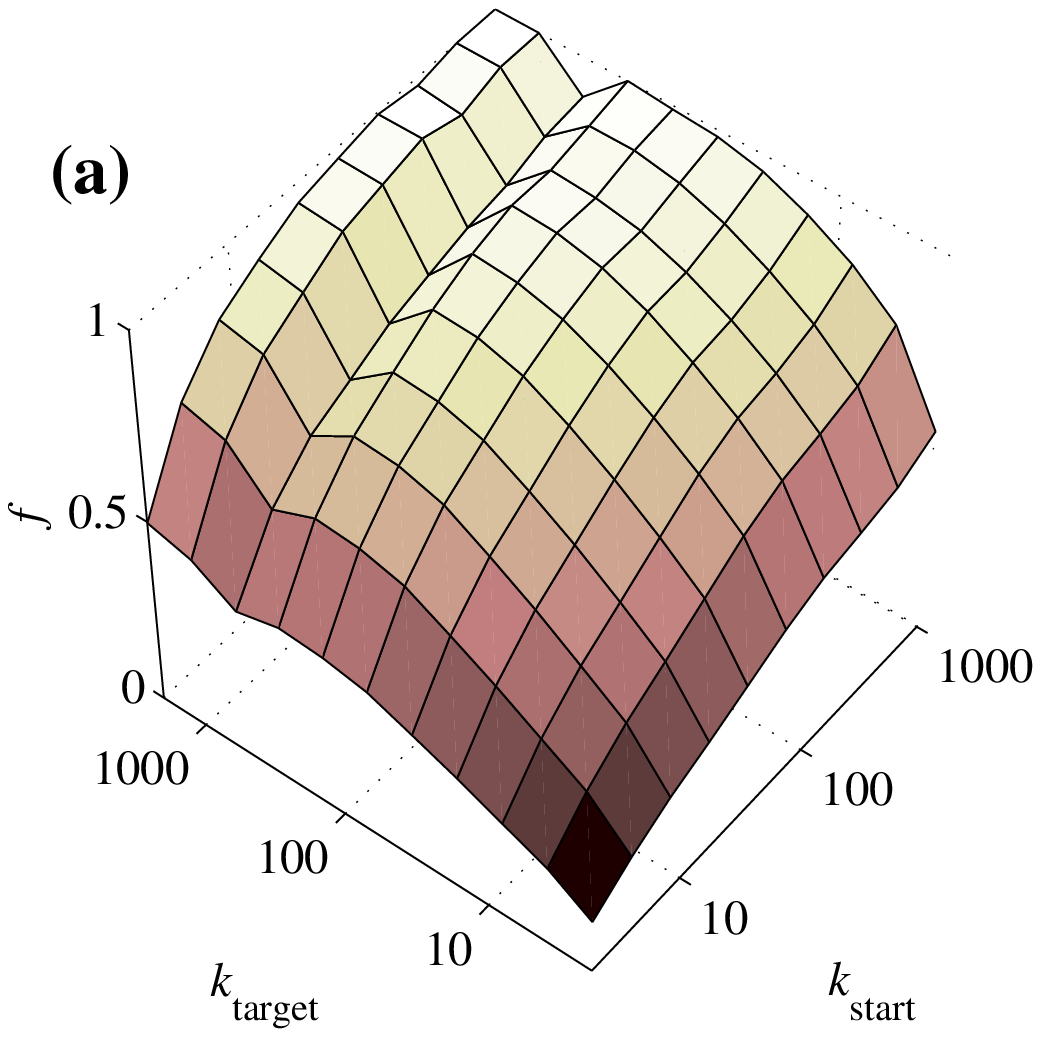}}
  \resizebox*{!}{6.5cm}{\includegraphics{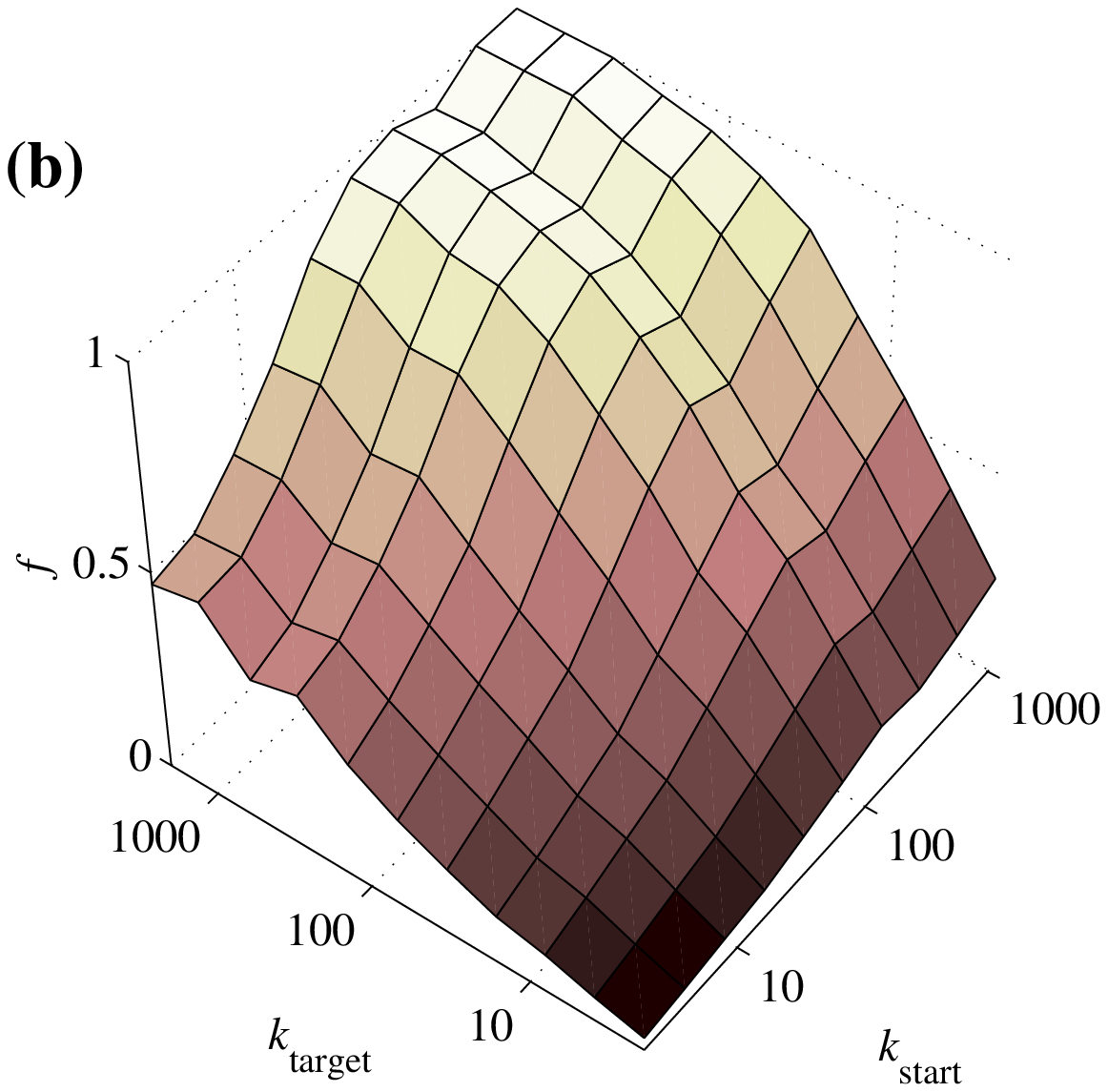}}\\
  \resizebox*{!}{6.5cm}{\includegraphics{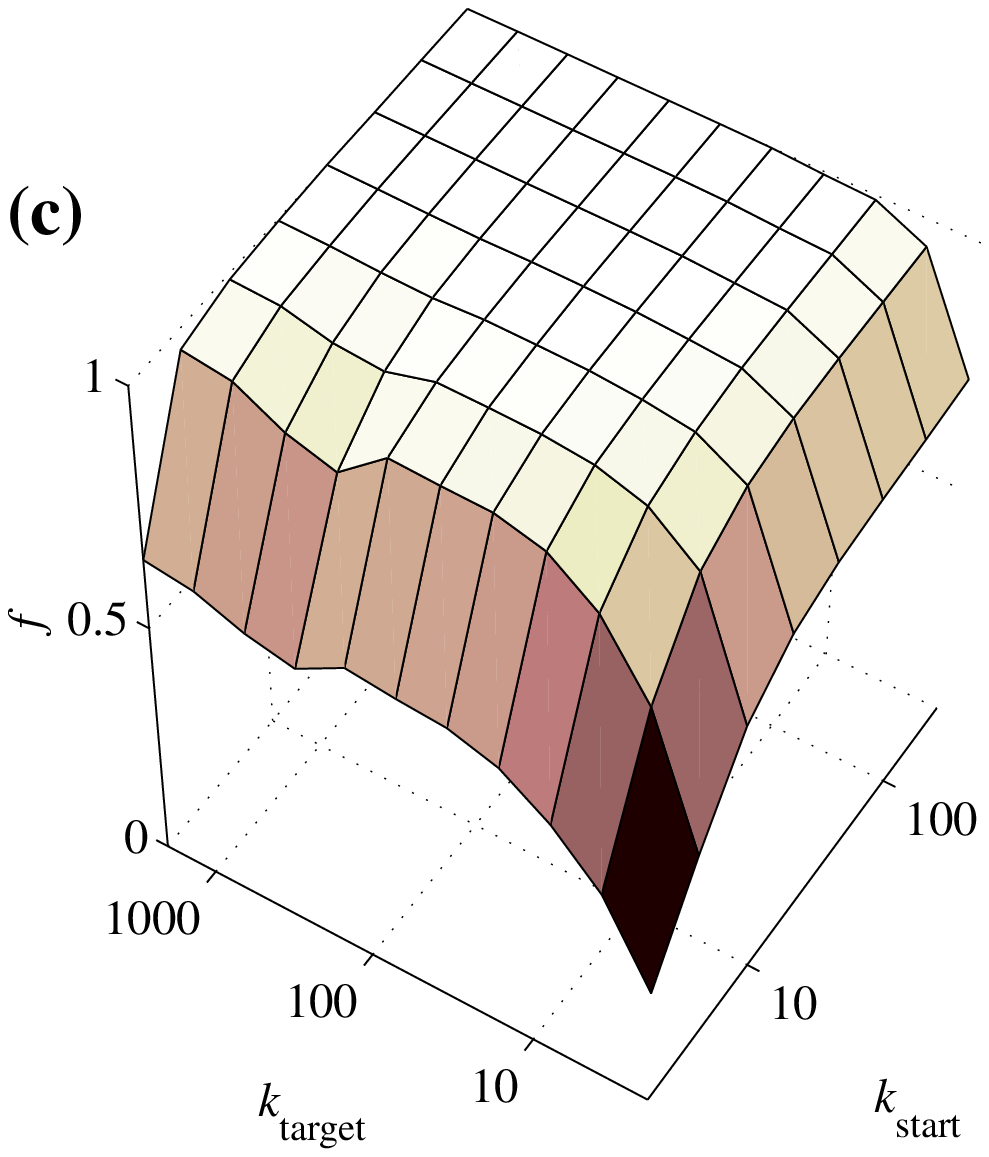}}
  \resizebox*{!}{6.5cm}{\includegraphics{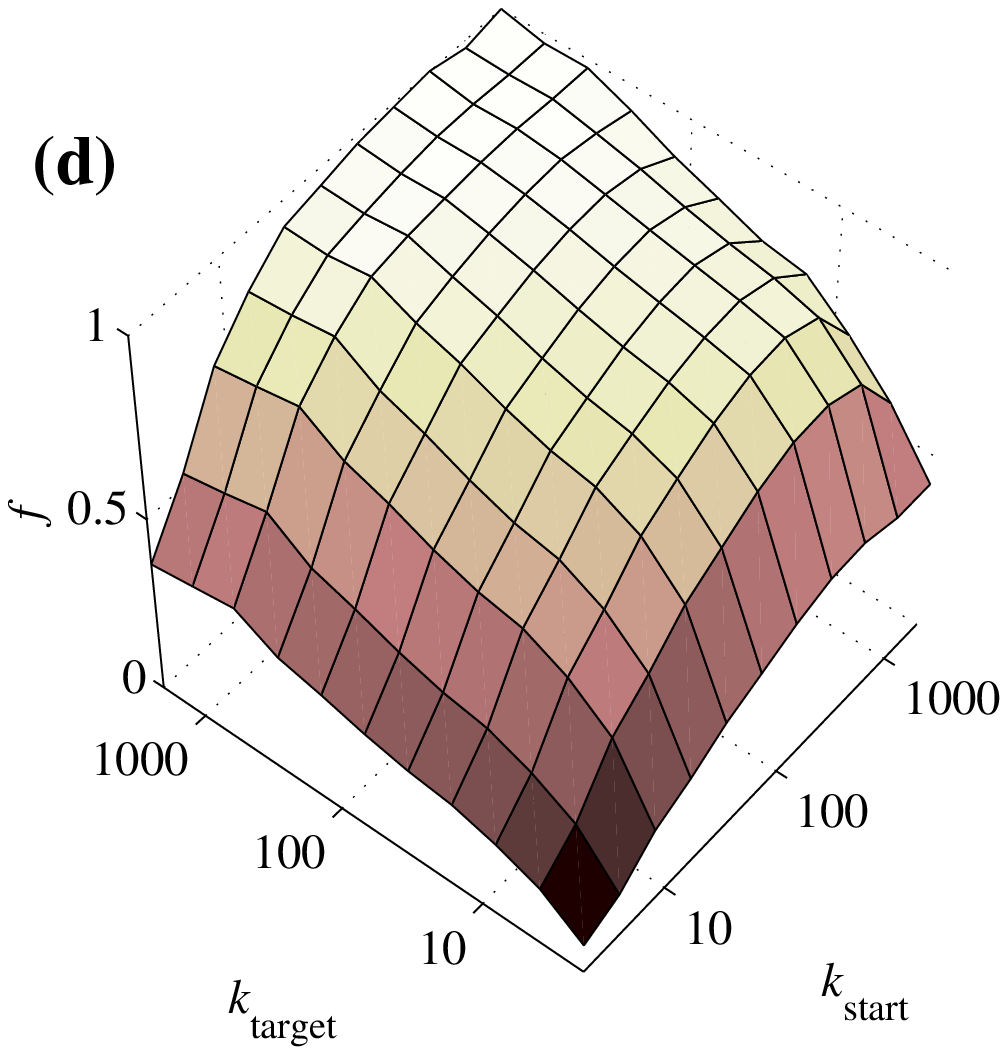}}
  \caption{ The reachability ratio $f$ as a function of the
    out-degree of the start vertex of the time respecting path
    $k_\mathrm{start}$ and the in-degree of the final vertex of the
    path vertex $k_\mathrm{target}$. The horizontal axes are
    logarithmically binned. (a) shows the result for the Internet
    community pussokram.com, (b), (c) and (d) shows the e-mail exchange
    sequences of Ebel \textit{et al.}, Eckmann \textit{et al.}\ and
    the Enron set respectively.
  }
  \label{fig:fkk}
\end{figure*}

\begin{figure}
  \resizebox*{\linewidth}{!}{\includegraphics{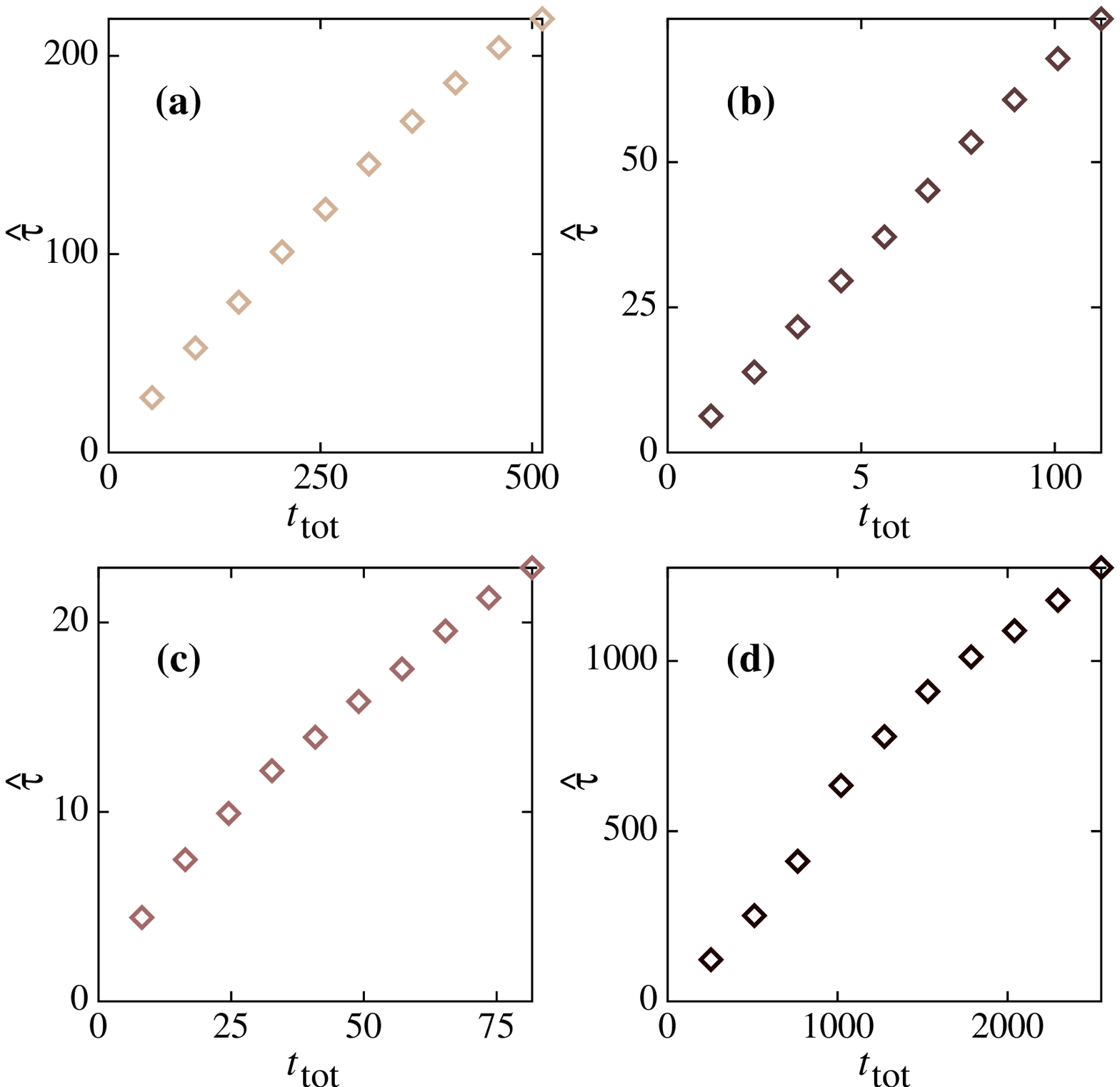}}
  \caption{
    The convergence of $\hat{\tau}$ as a function of the number
    sampling time of subsets of the contact sequences. The panes
    correspond to the different data set (with the same labeling as
    Fig.~\ref{fig:ncl}). The errorbars are smaller than the symbol
    size except for in (d) where they are of about the same size as
    the symbols.
} \label{fig:conv}
\end{figure}

Another way to extend RT to a more random ensemble, other than to
choose random times for the contacts along the edges such as in RC, is to
randomize the edges, but keep the set of degrees and the set of
contacts per edge fixed. This approach, common for static
networks~\cite{sni9ij}, results in unanimously longer
reachability times, and smaller reachability ratios, than for RT. The structure lost
by ER randomization (with respect to RT) is the degree-degree
correlations and clustering (a heightened probability for short
circuits). These networks have close to neutral degree-degree
correlations~\cite{my:ongoing}, and a slightly increased clustering
compared to ER randomized networks. One possible explanation for the
lower ER reachability is that the ER randomization removes a
positive $l$-correlation between adjacent edges. Supposing the
broad $l$ distribution of Fig.~\ref{fig:cldist} to some extent is due
to some persons being more active communicators than others, it is
likely that a large number of edges with many contacts lead to and
from such an individual, thus creating such an correlation. That there
really is a positive $l$-$l$-correlation for adjacent edges is shown
in Table~\ref{tab:llc}. This positive correlation means that the
passage, i.e.\ a paths of length two (a considerable distance since
these are all small-world networks with very short average distances),
through such a highly active communicator is very rapid. A perhaps even
more significant correlation removed by RE is the increased
communication rates for edges connecting vertices of high degree. This
is described for airline networks and networks of scientific
collaborations in Ref.~\cite{bar:wei} but holds for our networks too,
as shown in Fig.~\ref{fig:lkk}. Since our networks are directed, we plot $l$
as a function of the out-degree of the from-vertex (the edge's first
argument) and in-degree of the to-vertex (the out-degree is related to
the outward contact activity of the from-vertex and is therefore
be the relevant degree). We see very high communication along edges
connecting the vertices with highest degree for all our networks. (For
the Ebel \textit{et al.}\ data in Fig.~\ref{fig:lkk}(b) the highest
$l$ value is so high, $\sim 1000$, that we truncate the plot to be able to see the
rest of the structure, and to be able to compare with the other
networks. This is probably related to the edges with unrealistically
high communication rates discussed above.) Edges between vertices of
high degree are known to carry many shortest paths~\cite{our:attack},
this correlation means that the shortest paths, in the real data sets,
are possible to utilize in time respecting paths. Clearly the
removal of this structure by RE randomization will decrease the
reachability.

The AR randomized networks are Poisson random graphs with the same $N$
and $M$ as the original networks, and the $L$ contacts spread out with
uniform probability over the same interval as the original contact
sequences. For the three denser (in $k$ and $l$) data sets, the pussokram.com, Eckmann
\textit{et al.}\ and the Enron data, the AR scheme gives the highest
reachability ratio. For the very sparse Ebel \textit{et al.}\ data AR
lower $f$ considerably. The effect must partly have a dynamic
explanation as this data set is above the threshold for
a giant strongly connected component ($M=N$)~\cite{mejn:arb}.

\subsection{Reachability and the characteristics of individual
    vertices and edges}

So far we have dealt with average reachability properties. In this
section we will take a brief look on how the reachability depends on
the characteristics of the individuals. In Fig.~\ref{fig:fkk} we plot $f$
as a function of the out-degree $k_\mathrm{start}$ of the first vertex
of a time-respecting path and the in-degree $k_\mathrm{target}$ of the
last vertex of the path. The reason to look a the out-degree of the
start vertex and the in-degree of the target vertex is similar as for
the to- and from vertex in the context of Fig.~\ref{fig:lkk}---these
are the relevant degrees for the number of paths leading to and from
the vertices. We see that for all data sets the $f$ for medium degree
vertices is almost as high as for the vertices of highest degree, only the
low degree vertices have significantly lower reachability ratios---on
the other hand the low-degree vertices are, by far, the most frequent
in the data sets. Plots of $\hat{\tau}$ against $k_\mathrm{start}$
and $k_\mathrm{target}$ give the same picture---reachability in this
sense is also strictly increasing with the degree of the start- and
target vertices. The picture emerging is that the contact sequences
have a core \cite{chung_lu:pnas,addh} of not only a very small diameter but
also (cf.\ Fig.~\ref{fig:lkk}) very fast dynamics.

\subsection{Extrapolations to the large-time limit}

Statistical physics traditionally deals with the limit of
large size. In this section we discuss the relevance of
this limit and how it can be studied through finite size scaling. An
intuitive method is to average truncated intervals
$[t'_\mathrm{start},t'_\mathrm{stop}]\subseteq
[t_\mathrm{start},t_\mathrm{stop}]$ of increasing length. In
Fig.~\ref{fig:conv} we plot $\hat{\tau}$ for our four data sets as a
function of the sampling time averaged over 100 intervals of every
length. The curves are increasing with a convex shape, suggesting
either a convergence to a non-zero value or a divergence towards
infinity. We note that the $t\rightarrow\infty$ limits of both
$\hat{\tau}$ and $f$ are non-trivial---there can of course be time
respecting paths of the order of the sampling time, but the fraction of
such long paths will decrease. Thus the convexity of the
$\hat{\tau}(t)$-curves is quite expected. There are many time-scales present
in the data: Above we mention days and weeks that separate recurring
routines. The average duration of an edge was studied in
Ref.~\cite{my:ongoing}. Furthermore the average time between contacts
per vertex, and the life-time of vertices both define time
scales. The finite size scaling method discussed above is useful
for processes affected by time-scales similar to, or shorter than, the
sampling time. It is hard to use it to deduce the $t\rightarrow\infty$
limits of $\hat{\tau}$ and $f$. So in the absence of data sets with
considerably longer sampling times, we have to focus on questions such
as how the contact patterns affect the network reachability (discussed
in the previous two sections) and short-time dynamic processes.

\section{Summary and discussion}

We have studied how fast information can spread among people involved in electronic
communication. Four data sets---one from communication within an
Internet community and three e-mail data sets---are used. We find that these data
sets can be characterized by a core with short path lengths and
frequent contacts and a periphery to which information is
relatively unlikely to reach. We note that the distribution of
contacts per edge is highly skewed, and if this was not the case, the
dynamics would be much faster.  The fact that people engage in dialogs
with others tend to slow down the dynamics (compared to if the sending
times were independent of earlier communication). This implies that
the order of the contacts matter much, and one loose the full
picture by converting contact sequences to weighted networks. 
We discuss finite size scaling by truncating the sampling interval and
conclude that it is very useful method, but only for processes of the
same time scale, or faster, as the sampling time.

In the discussion of the data sets, we mention the e-mail data sets
are incomplete with respect to outer vertices (vertices not belonging
to the sampled e-mail server for the Ebel \textit{et al.}\ and Eckmann
\textit{et al.}\ data) or not belonging to the sampled e-mail
directories (the Enron data). E-mails between outer vertices are not
recorded in the Ebel \textit{et al.}\ data and only partially present
in the Enron data whereas the Eckmann \textit{et al.}\ data does not
include messages to outer vertices. We note that, while these
different sampling procedures certainly affects many quantities (see
Fig.~\ref{fig:ncl}), the conclusions above about how different
structures affects the reachability remains intact.

Not all messages contain information that of the kind
that people would relay to others. Technically, considering also the
finite size of the sampling intervals, we obtain lower bounds
on the reachability time and upper bounds on the reachability
fraction. On the other hand we believe that our qualitative
conclusions would be unaltered for the averages of $f$ and
$\hat{\tau}$ in real information spreading.

So far, the spreading processes we have mentioned has been the
diffusion of rumors and similar information. Our results may also
apply to computer virus epidemiology. There is, presumably, a big
difference between the contact sequences of viral and regular emails
as the former often are sent to many recipients in a short time period, but
only once per recipient~\cite{mejn:liame}. The methods we use, however,
are be perfectly applicable to a sequence of computer virus transmissions.

There are many ways to extend this work. For example not only the
shortest time respecting paths are relevant for spreading dynamics of
different kinds---information might spread by a longer time
respecting path. A sensible extension of this study would therefore to
include these longer paths in the picture (cf.\
Ref.~\cite{flow:betw}). Much work remains to get a full statistical
characterization of contact sequences, not to mention the
understanding of how dynamical systems of e.g.\ epidemiology are
affected by by dynamical properties of the contact patterns.

\subsection*{Acknowledgments}

We thank Holger Ebel and Jean-Pierre Eckmann for help with the data
acquisition and Yong-Yeol Ahn, Michael Gastner, Mark Newman and Juyong Park for
comments.

\end{document}